# Doping dependence of the critical fluctuation regime in the Fe-based superconductor $Ba_{1-x}K_xFe_2As_2$

Jianqiang Hou[1#], Philipp Burger[2#], Huen Kit Mak[1], Frédéric Hardy[2], Thomas Wolf[2], Christoph Meingast[2] and Rolf Lortz[1♠]

[1] *Department of Physics, The Hong Kong University of Science and Technology, Clear Water Bay, Kowloon, Hong Kong, China*

[2] *Institute for Solid State Physics, Karlsruhe Institute for Technology,, PO Box 3640, 76021 Karlsruhe, Germany*

We investigate the importance of superconducting order parameter fluctuations in the 122 family of Fe-based superconductors, using high-resolution specific heat and thermal expansion data of various $Ba_{1-x}K_xFe_2As_2$ single crystals covering a large range of the phase diagram from the strongly underdoped to the overdoped regime. By applying scaling relations of the 3d-XY and the 3d-Lowest-Landau-Level (3d-LLL) fluctuation models to data measured in different magnetic fields, we demonstrate that a strong increase of the critical fluctuation regime is responsible for the transition broadening in magnetic fields, which is a direct consequence of a magnetic-field-induced finite size effect due to a reduction of the effective dimensionality by a decreasing magnetic length scale related to the mean vortex separation and the confinement of quasiparticles in low Landau levels. The fluctuations are stronger in the underdoped and overdoped regimes and appear to be weakest at optimal doping.

## I. INTRODUCTION

Superconductors are characterized by a two-component order parameter, in which the phase and amplitude of the Cooper pair wave function represent the two degrees of freedom. In cuprate high-temperature superconductors with their high transition temperatures, the low-dimensional layered structure and the associated small coherence volumes, critical order parameter fluctuations have a strong impact on their superconducting properties in a large temperature regime around their critical temperature [1]. For, example for the case of optimally and overdoped $YBa_2Cu_3O_{7-\delta}$, physical quantities, such as the superconducting correlation and coherence lengths [2], specific heat [3] and thermal expansion [4,5], follow power law dependencies with critical exponents of the 3d-XY universality class. The superconducting transition thus belongs to the same universality class as the superfluid transition in liquid helium

[#] These authors contributed equally to the article.

[♠] Corresponding author: lortz@ust.h

[6], and, consequently, the specific heat and thermal expansion anomalies show the characteristic lambda-shape anomaly [2,3,7]. In classical superconductors, the low critical temperatures and large coherence volumes limit the critical fluctuation regime to an extremely small temperature interval, which is typically experimentally inaccessible. However, a strong applied magnetic field introduces a magnetic length scale that is associated with the mean vortex-vortex separation and thus imposes a spatial limit to the coherence length in the lateral plane [8]. In addition, the magnetic field confines the quasiparticles in low Landau levels. As a result, the effect of fluctuations can become experimentally observable, which has been shown for example for the classical superconductors $Nb_3Sn$ [9] and $V_3Si$ [10]. Signatures of critical fluctuations were also observed in some iron-based superconductors. They appear to be stronger in the more anisotropic 1111 compounds [11] than in the 122 compounds [12,13] and their overall strength appears to be between the behaviors in the cuprates [2] and the classical superconductors [9,10].

In this paper, we investigate the specific heat and thermal expansion of various single crystals of $Ba_{1-x}K_xFe_2As_2$ in magnetic fields with x = 0.22, 0.35, 0.50 and 0.60. We have previously reported the observation of a vortex melting transition in the specific heat for the slightly overdoped single crystal with x = 0.5 [14], which unambiguously demonstrated the high quality of the present samples. Here, we show that the data measured in different magnetic fields follows the scaling predicted by the 3d-XY and 3d-LLL fluctuation models and demonstrate that, similar to $YBa_2Cu_3O_{7-\delta}$ [2], $Nb_3Sn$ [9] and $V_3Si$ [10], the superconducting transition in large magnetic fields is rendered into a continuous crossover, induced by a magnetic field induced finite-size effect [2]. The critical regime is larger in the underdoped and overdoped regimes of the phase diagram and has a minimum at optimal doping.

## II. EXPERIMENTAL

$Ba_{1-x}K_xFe_2As_2$ single crystals were grown from self-flux in an $Al_2O_3$ crucible. Ba and K were mixed with prereacted FeAs in the desired ratio and filled into the crucible. The crucible was sealed in a steel container and heated to 1151°C. The crucible was then cooled down very slowly to 1051°C at 0.2 – 0.5 °C/h. In order to decant the remaining flux, the crucible was tilted at the end of the growth process and slowly pulled out of the furnace. The slow cooling rate allowed us to grow large single crystals of high homogeneity with 4 - 8 mm length in the *ab* plane, which is crucial for high resolution thermal expansion measurement. The exact K content *x* of the samples typically differs from the starting stoichiometry, and was determined by energy-dispersive x-ray analysis (EDX) and by 4-circle x-ray diffraction.

The specific-heat experiments were performed under field-cooled conditions with a home-made high-resolution micro-calorimeter [14]. The micro-calorimeter allowed us to measure a very tiny 60 µg crystal with a particularly high homogeneity and thus sharp $T_c$. It can be used with either a DC 'long-relaxation' technique for measurements with precision up to 1%, or with a modulated-temperature AC technique. The latter is less accurate, but provides higher resolution of $\Delta C/C=10^{-5}$ and a high data point density. The data shown in this paper was measured with the AC mode, but the absolute values have been calibrated with the relaxation mode. The high resolution thermal expansion measurements were performed with a capacitive dilatometer, which offers resolutions of $\Delta L(T) = 10^{-2}$ Å for absolute length changes of the single crystals. The linear thermal expansion coefficient is defined as the partial derivative of the crystal length along a certain crystallographic direction $i$ by the temperature $\alpha_i(T) = \frac{1}{L_i}\frac{\partial L_i}{\partial T}\Big|_p$. In this article we focus on the crystalline $a$-axis, for which the largest superconducting anomalies are found in $Ba_{1-x}K_xFe_2As_2$. In the vicinity of a second-order phase transition, the specific heat and the thermal expansion coefficient are closely related via the thermodynamic Ehrenfest relation $\frac{\partial T_c}{\partial p_i}\Big|_{p_i \to 0} = V_{mol} T_c \frac{\Delta \alpha_i}{\Delta C_p}$ [15], (where $\partial T_c/\partial p|_{pi\to 0}$ is the uniaxial pressure dependence of $T_c$ in the limit of zero pressure, $p$ is the pressure and $V_{mol}$ is the molar volume of the material). Therefore, the anomalies at the superconducting transition will have the same shape and can be analyzed in a similar manner. In contrast to the anomaly $\Delta C_p(T=T_c)$, which always has a positive signature, the signature of $\Delta\alpha(T=T_c)$ can have a positive or negative signature, dependent on the signature of $\partial T_c/\partial p_i$.

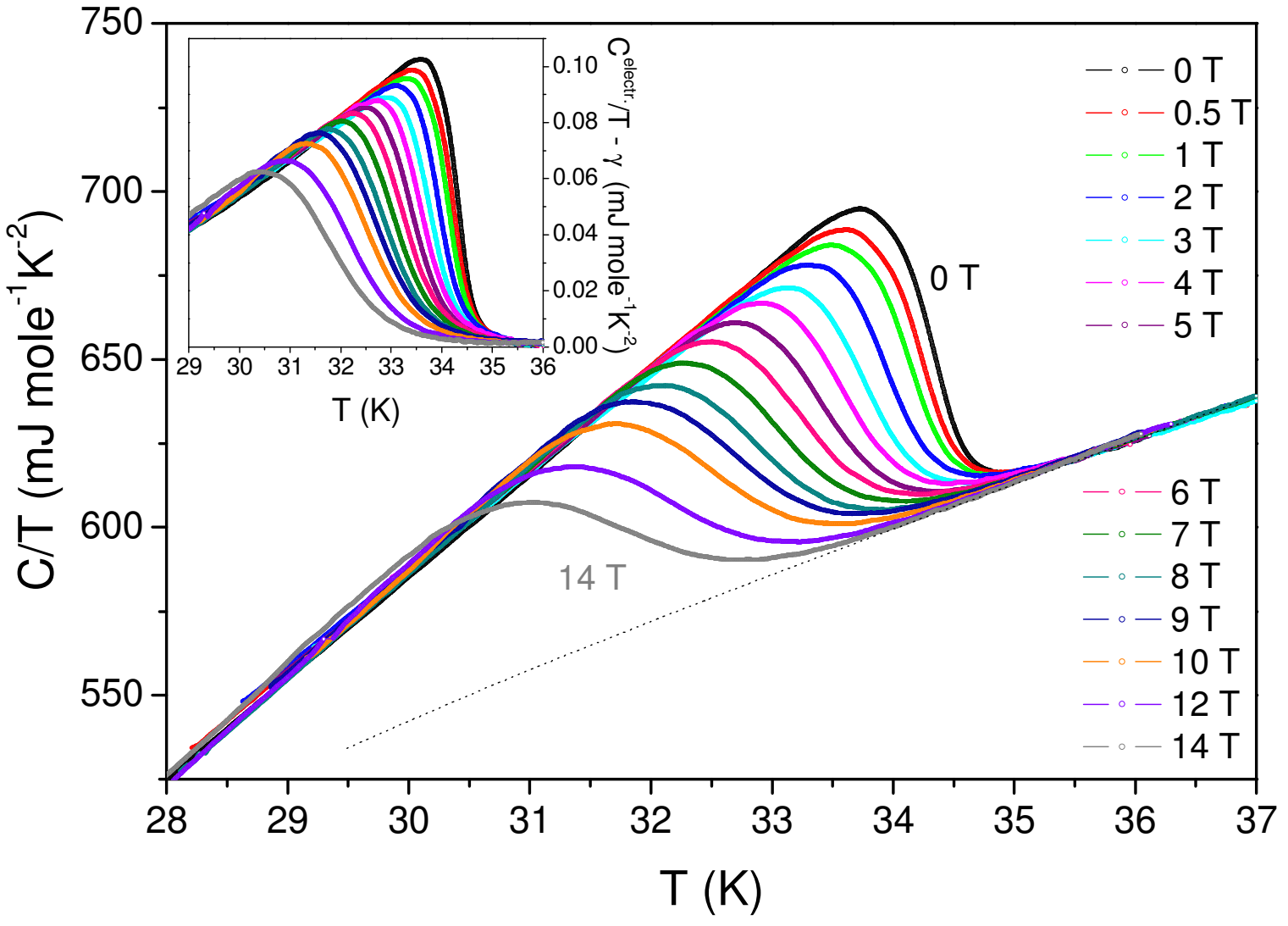

**Figure 1.** Total specific heat of $Ba_{0.5}K_{0.5}Fe_2As_2$ in various magnetic fields up to 14 T applied parallel to the crystalline $c$-direction. The dotted line shows an estimation of the normal state background contribution. Inset: jump at the superconducting transition in $C^{electr.}(H) - \gamma$, after separation of the phonon contribution and the normal state Sommerfeld constant $\gamma$.

## III. RESULTS

Fig. 1 shows the total specific heat of a single crystal with x = 0.5 for magnetic fields between 0 and 14 T applied parallel to the crystalline $c$-axis. The relatively sharp specific-heat jump occurs in zero field at $T_c$ = 34.3 K and is lowered in applied fields down to 32.7 K in 14 T. A strong magnetic-field-induced broadening of the transition is obviously observed in the data of Fig. 1, and, as we show later, is a fingerprint of the growing strength of fluctuations. We have previously shown by subtracting a suitable background that further tiny steps and peak-like anomalies occur at temperatures slightly below the broadened superconducting transition due to a vortex melting transition [14]. The melting transition line in the magnetic field versus temperature phase diagram was found to follow a power law dependence of $B_m \sim [1-T_m/T_c(0T)]^{0.67}$. A similar characteristic power law dependence has been observed for the vortex melting transition in $YBa_2Cu_3O_{7-\delta}$ and has been interpreted in the framework of phase fluctuations of the 3D-XY universality class with a critical exponent $\nu = 0.669$ [16,17].

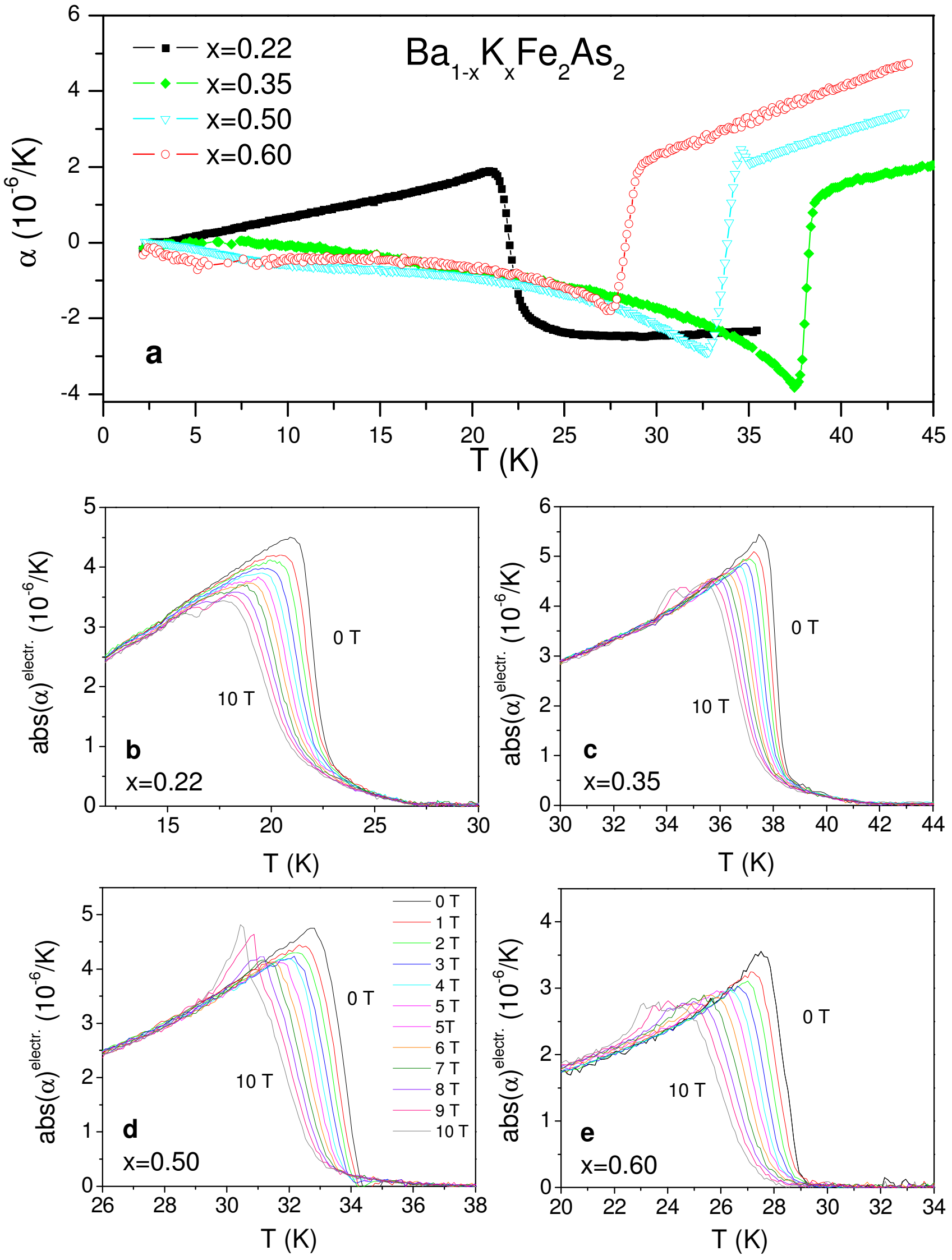

**Figure 2. a)** zero-field linear thermal expansion coefficient measured along the crystallographic a-axis of 4 single crystals of $Ba_{1-x}K_xFe_2As_2$ with x = 0.22, 0.35, 0.50 and 0.60. **b – e:** Electronic contribution of the 4 single crystals in magnetic fields of 0, 1, 2, 3, 4, 5, 6, 7, 8, 9 and 10 T in absolute values after separation of the phonon and electronic normal-state contributions.

Fig. 2a shows the linear thermal expansion coefficient $\alpha(T)$ along the crystallographic $a$-axis for $Ba_{1-x}K_xFe_2As_2$ with x = 0.22, 0.35, 0.50 and 0.60 [18]. The superconducting transition appears as a jump in $\alpha(T)$, which is of negative signature for x = 0.35, 0.50 and 0.60 and of positive signature for x = 0.22. The different sign for the latter results from a competition between SDW state and superconductivity [19,20]. The data presented here were measured upon heating the sample with a magnetic field applied parallel to the crystalline $c$-axis after cooling in the same field. In addition to the main superconducting anomalies, sharp peaks appear below $T_c$ in the vicinity of the vortex melting temperature. We have previously shown that these peaks are related to an irreversible component of the thermal expansion and appear only upon heating the sample [14]. They are caused by flux gradients that cause non-equilibrium screening currents to form during cooling of the samples. These screening currents apply a force on the sample via the pinning centers. When approaching the vortex melting transition or the irreversibility line, they decay rapidly, thus causing a sharp peak in $\alpha(T)$. We decided to use data in this analysis measured upon heating, because the peaks also provide information about the scaling behavior of the irreversibility or vortex melting transition line.

In the inset of Fig. 1, as well as in Fig. 2b – 2f, we show the superconducting component of the specific heat and the linear thermal expansion coefficient for the different samples, respectively, after subtraction of an estimated normal-state background. The latter was obtained by a smooth polynomial extrapolation. Note, that the ratio of the $T_c$ anomaly to the phonon background is particularly large in the thermal expansion, so that the exact choice of the background has little influence on the data analysis, which was tested carefully. The thermal expansion data allows us to compare the shape of the transition and its field dependence over a large doping range from strongly underdoped (x = 0.22) to overdoped (x = 0.60). A closer look reveals a significant change of behavior upon increasing K content. While fluctuation components appear for the underdoped to optimally doped samples with x = 0.22 and 0.35 in form of a high-temperature tail above the transition (for example up to ~41 K for x = 0.35, 3 K above the midpoint of the jump in $\alpha(T)$), this tail is largely absent in the overdoped samples with x = 0.5 and 0.6. Although there is some uncertainty in the exact normal-state background, this trend can already be seen in the original data and does not originate from improper background estimation. Especially for the sample with x = 0.6 it is obvious in the raw data that the data follows a linear behavior just above the jump in the thermal expansion with almost not fluctuation contribution, while the most underdoped sample shows fluctuations up to 3 K above $T_c$. Another difference is seen in the field dependence. The applied magnetic field has little effect on the high temperature fluctuation tail in the underdoped samples and basically only broadens the phase transition, thus lowering the temperature at which the maximum of the anomaly appears. The situation is different in the overdoped samples, where the onset temperature of the transition is clearly decreased by applied fields. The overall behavior for the underdoped samples is similar to $YBa_2Cu_3O_{7-\delta}$, in which the transition is strongly subject to 3d-XY fluctuations [2]. The overdoped samples show a more BCS-like transition behavior, although fluctuation components are present at lower temperatures, as we will show later.

## IV. DISCUSSION

If the observed broadening of the superconducting transition in applied magnetic fields is caused by the increasing strength of fluctuations, then it should be possible to scale all the magnetic field curves onto a single scaling curve using an appropriate fluctuation theory [21-28]. Such scaling of specific heat or thermal expansion data in various applied fields has been previously applied to cuprate high temperature superconductors [2,9,24-28] and the classical superconductors $Nb_3Sn$ [9] and $V_3Si$ [10]. The magnetic field introduces a magnetic length $(\Phi_0/B)^{1/2}$, which reduces the effective dimensionality and thus imposes a limit to the correlation volume [2,8], which would otherwise diverge at $T_c$. For the scaling, data taken in different fields are normalized by the ratio of the coherence volume to the magnetic length. If fluctuations dominate the thermal properties of the sample in a regime around the critical temperature, scaled data should merge to a common curve. The 3D-XY universality class is strictly valid only in the limit of small fields, although it has been shown for $YBa_2Cu_3O_{7-\delta}$ that this range can span several Teslas. In higher fields, the quasiparticles are confined in low Landau levels that induce a dimensional crossover towards 1D. Therefore, the 3d-LLL approximation (where LLL stands for 'Lowest Landau Level') should be applicable, which predicts a slightly different scaling behavior [23,24]. There is no general consensus, in which magnetic field range this crossover should occur. It has been observed to occur in $Nb_3Sn$ in a field of ~1 T [9], but for $YBa_2Cu_3O_{7-\delta}$ it was not observed in the field range up to 11 T [2].

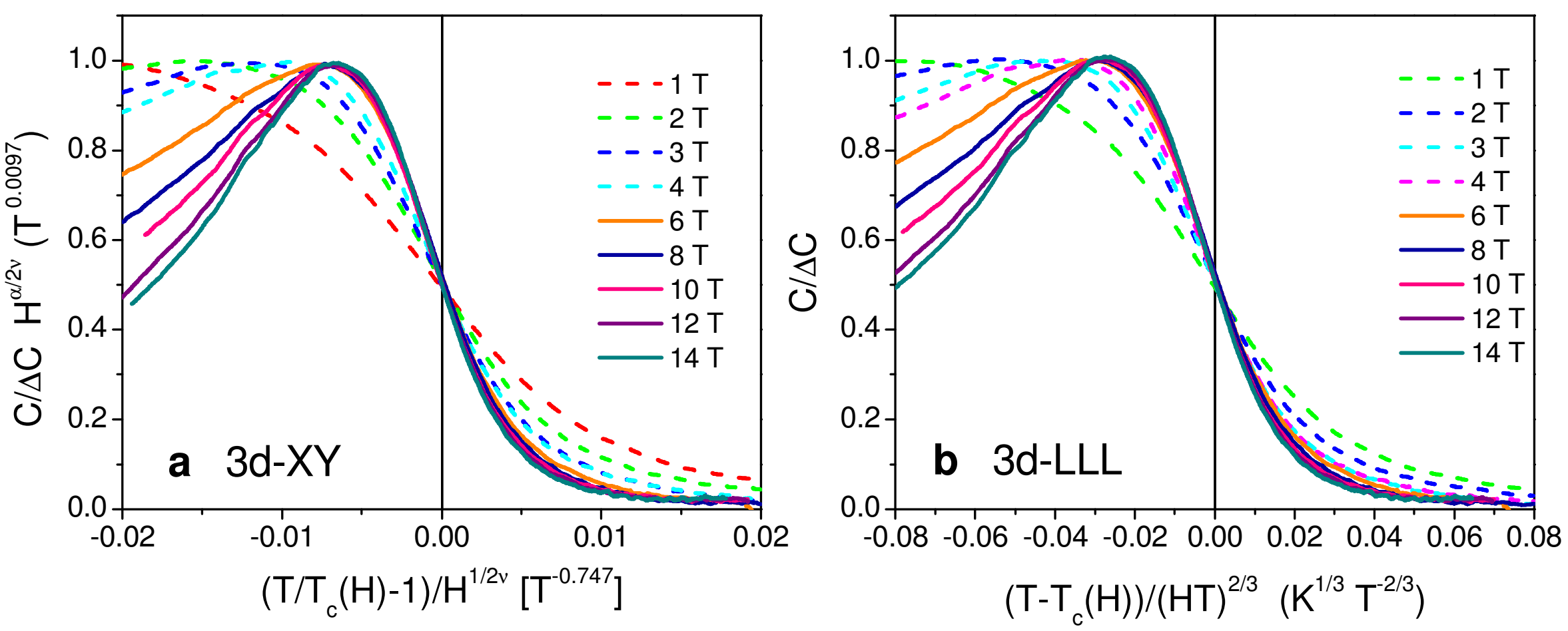

**Figure 3. a** 3d-XY and **b** 3d-LLL scaling of the specific heat of $Ba_{0.5}K_{0.5}Fe_2As_2$ in various applied magnetic fields after subtraction of the normal-state contribution. Scaling is observed for both models in fields exceeding 4 T.

In Fig. 3a and 3b we have plotted the electronic specific heat of $Ba_{0.5}K_{0.5}Fe_2As_2$ near $T_c(H)$ as a function of the 3d-XY and 3d-LLL scaling variables, respectively. Since the fluctuation contribution is small, we normalized the mean-field specific heat jump component and considered a field-dependent $T_c(H)$ [9]. If critical 3d-XY fluctuations are present, the data should merge if plotted as $C/\Delta CH^{\alpha/2\nu}$ versus $[T/T_c(H)\text{-}1]H^{-1/2\nu}$ ($\nu \cong 0.669$, $\alpha \cong -0.007$)) [21]. The scaling fails for fields below 4 T. This is not surprising, since the zero-field transition is already broadened by chemical inhomogeneity. In higher fields, where the field-induced broadening exceeds the 'chemical broadening', the scaling improves rapidly and holds well in the field range of 4 T and above. The scaling is limited to the temperature range between the maximum of the transition anomaly and the upper onset, which indicates the width of the critical regime. At lower temperatures the curves start to diverge, which is a natural consequence of the decreasing strength of fluctuations and the resulting crossover to a mean field behavior and also observed in $YBa_2Cu_3O_{7-\delta}$ [2]. In Fig. 3b, we plotted the same data as $C/\Delta C$ versus $[T-T_c(H)](HT)^{-2/3}$, which is the scaling of the 3d-LLL model. Similar to the 3d-XY scaling, the 3d-LLL scaling works well in fields above 4 T in the region of the broadened specific-heat jump, while the curves differ in the lower field and temperature ranges. The similarity of the scaling diagrams of the two models is not surprising, since the exponents in the scaling variables of both models are almost identical. This makes it difficult to distinguish them (especially in the high field regime) while differences would appear in smaller fields where no scaling is observed in any case. For $YBa_2Cu_3O_{7-\delta}$ it was possible to rule out 3d-LLL fluctuations up to fields of 11 T only because of the much wider fluctuation range [2]. The need for introducing a field-dependent $T_c$, which is not included in the 3d-XY model, may mean that the data in fields exceeding 4 T fall in the 3d-LLL regime. However, it is more likely that the reason is that the fluctuations are weak and appear only as a small correction to a BCS-type mean field behavior. In any case, our main focus here is not to distinguish the 3d-XY and 3d-LLL scaling regimes, but rather to demonstrate the importance of fluctuations in $Ba_{1-x}K_xFe_2As_2$ and to explain the field-induced broadening of the transition by their presence. We can conclude from our specific heat data that the effect of fluctuations dominates in $Ba_{0.5}K_{0.5}Fe_2As_2$ in fields of 6 T and higher, while in lower fields the shrinking fluctuation regime becomes gradually hidden by the broadening of the transition due to the sample inhomogeneity.

In Fig. 4 we present the same scaling analysis performed for the linear thermal expansion coefficient $\alpha(T)$ for all available K concentrations. The result of the analysis is very similar to the specific heat. This is not surprising since both quantities are closely related. Both models offer a good scaling result for fields of 3 T and larger. The scaled data further confirm that the shape of the transition anomalies varies over the doping range. Samples with higher K-content show less rounded fluctuation contributions at higher temperatures. In cuprate high temperature superconductors, a doping-induced crossover from a 3d-XY fluctuation dominated behavior in the underdoped regime of the phase diagram to a more BCS-like mean-field behavior in the overdoped regime is observed [5], which appears to be similar at least for the underdoped regime of $Ba_{1-x}K_xFe_2As_2$. We will analyze this doping dependence more quantitatively later.

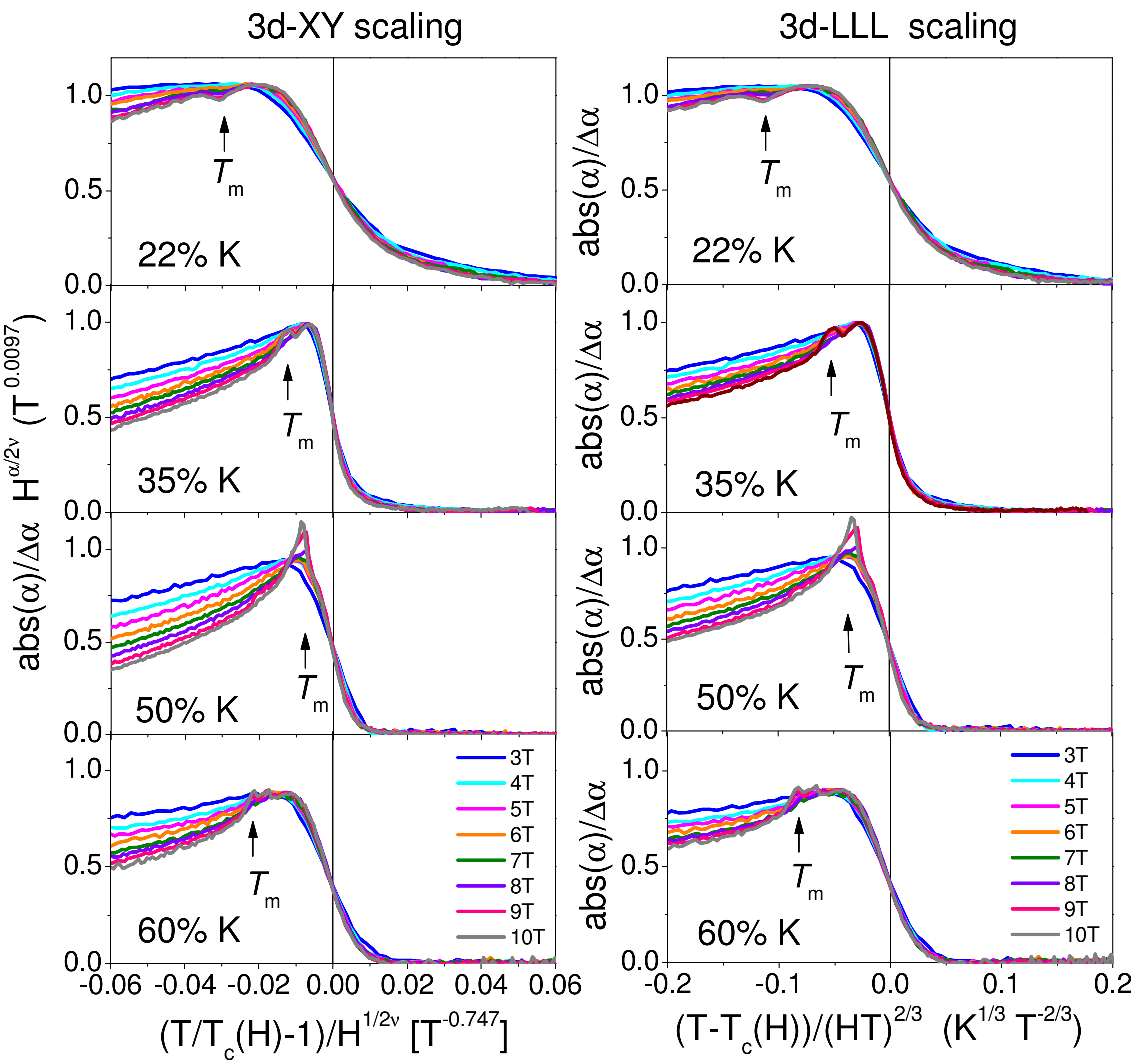

**Figure 4.** 3d-XY (panels in the left column) and 3d-LLL (right column) scaling of the linear thermal expansion coefficient $\alpha(T)$ along the crystalline *a*-axis of $Ba_{1-x}K_xFe_2As_2$ with x = 0.22, 0.35, 0.50 and 0.60 in various applied magnetic fields after subtraction of the normal-state contribution. The arrows mark irreversible anomalies that appear in the vicinity of the vortex melting transition (see text for details).

The observed scaling of the specific heat and thermal expansion coefficient suggests that the strong broadening of the specific heat anomalies in larger magnetic fields is caused by the increasing strength of superconducting fluctuations. One explanation for the magnetic-field-induced increase in the critical fluctuation regime can be found in the magnetic length $(\Phi_0/B)^{1/2}$ [2,8]. The Ginzburg temperature $\tau_G = Gi \cdot T_c$ marks the temperature range around $T_c$ where fluctuation contributions in the specific heat are of comparable magnitude as the mean-field jump ($Gi=0.5(k_B T_c)^2/(H_c^2(0)\xi_0^3)^2$: Ginzburg number, $H_c(0)$: thermodynamic critical field at $T$=0, $\xi_0$: isotropic Ginzburg-Landau coherence length). A reduction of the coherence volume by the limitation imposed by the magnetic length thus increases the Ginzburg temperature and will expand the critical regime [2]. This causes the observed strong broadening of the superconducting transition in magnetic fields. In higher fields, a further reduction of the dimensionality is imposed by the confinement of the quasiparticles in low Landau levels, which finally drives the crossover to the high-field 3d-LLL fluctuation regime [25]. Where the crossover from 3d-XY fluctuations to 3d-LLL fluctuations occurs in $Ba_{1-x}K_xFe_2As_2$ cannot be extracted from our data, since both scaling models can be applied equally well. This is surprising, since for $YBa_2Cu_3O_{7-\delta}$ [2] and $Nb_3Sn$ [9] no magnetic field range where both scaling models could be applied in parallel was found. Eventually, the crossover is particularly smooth in $Ba_{1-x}K_xFe_2As_2$, which may be a consequence of its intermediate fluctuation strength between cuprates and classical superconductors.

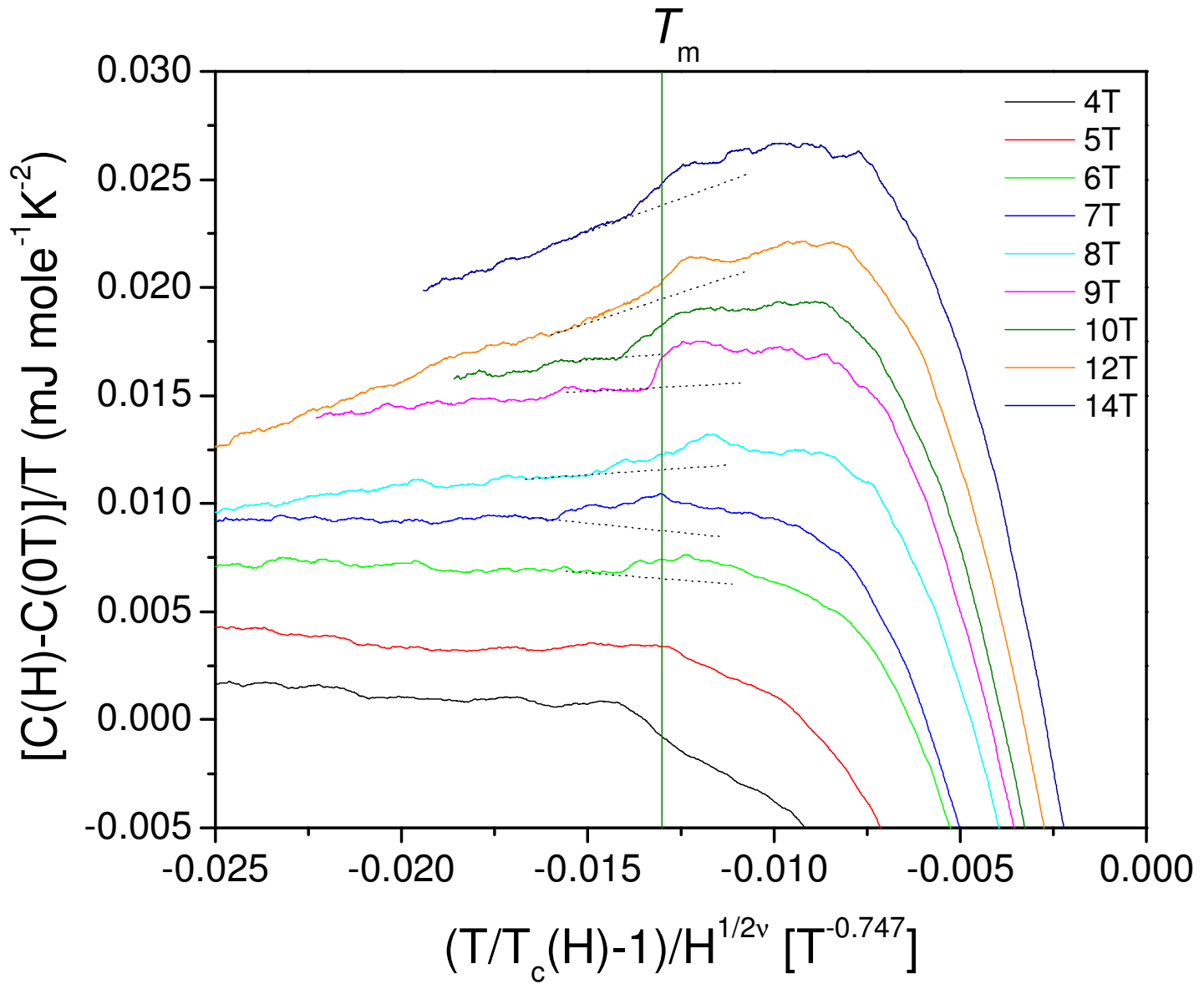

**Figure 5.** Specific heat of $Ba_{0.5}K_{0.5}Fe_2As_2$ as a function of the 3d-XY scaling variable $[T/T_c(H)\text{-}1]H^{-1/2\nu}$ in various applied magnetic fields with the zero field data subtracted as a background. In this plot the tiny anomalies at the vortex melting transition become visible [14], which occur at a fixed value of the scaling variable, as illustrated by the vertical line. The downturn on the right hand side is due to the subtracted zero-field superconducting transition anomaly. Except for the 4 T data, offsets of 2.5 mJ mole$^{-1}$K$^{-2}$ have been added for clarity.

In Fig. 5 we plot the same specific heat data once more as a function of the 3d-XY scaling variable $[T/T_c(H)\text{-}1]H^{-1/2\nu}$, but this time after subtraction of the zero field data as a background. This removes the steep background below $T_c(H)$ and makes the tiny vortex melting anomalies visible, which we have discussed previously in detail in Ref 14. The scaling plot demonstrates that the vortex melting transition occurs at a fixed value of $[T_m/T_c(H)\text{-}1]H^{-1/2\nu}$. This shows that the melting transition occurs when the order parameter fluctuations reach certain strength.

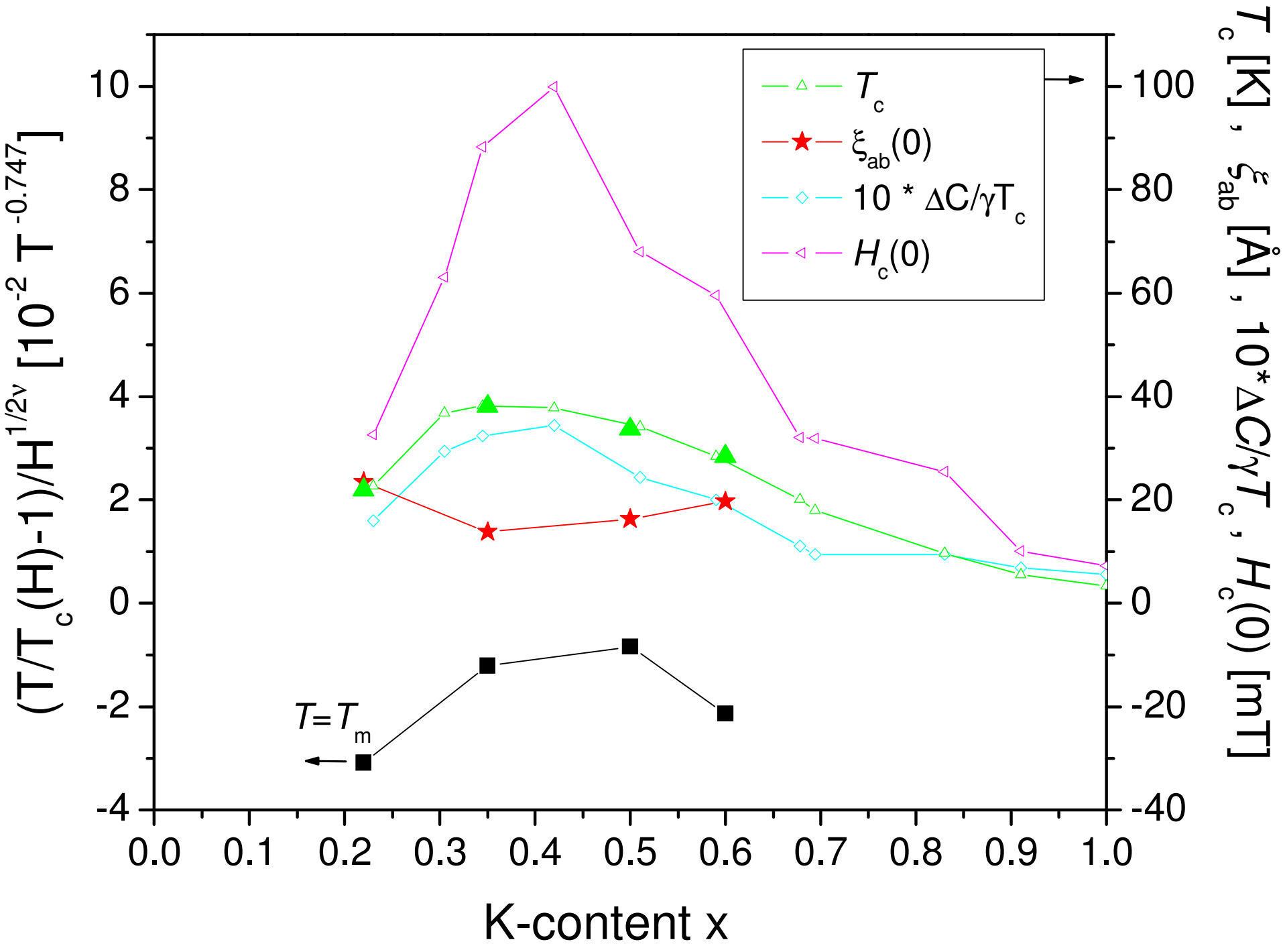

**Figure 6.** 3d-XY scaling variable where the vortex melting anomalies at $T_m$ occur in Fig. 4 as a function of K content x. Its absolute value represents a measure of the fluctuation strength. In addition, the green triangles show the corresponding $T_c$ values (the open triangles are added from Ref. 30), the red stars the doping dependence of the in-plane coherence length $\xi_{ab}(0)$ obtained for the same samples [18], the cyan diamonds are the normalized specific heat jump $\Delta C/\gamma T_c$ [30], and the magenta tilted triangles are the thermodynamic critical field $H_c(0)$ [30].

This scaling is also shown by the additional irreversible anomalies in the thermal expansion coefficient (Fig. 4), which are related to the vanishing of flux pinning in the vicinity of the vortex melting transition [16]. For all samples, the anomalies are found to follow the scaling predictions. However, for the different samples the anomalies appear at different fixed values of the x-scaling variable. The value of the x-scaling variable in the 3d-XY model can thus serve as a precise estimation of the approximate lower boundary of the fluctuation regime[1], and thus of the width of the fluctuation regime below $T_c$. Fig. 6 shows the characteristic value of the 3d-XY x-scaling variable where vortex melting occurs as a function of K-content, thus illustrating the doping dependence of the fluctuation regime width below $T_c$ at a fixed magnetic field value. It is obvious, that the width of the fluctuation regime is largest in underdoped samples and reaches a minimum at optimal doping before it increases again for the overdoped sample with x = 0.6.

[1] Note that we could base our argumentation in a similar manner on the 3d-LLL scaling variable, since both scaling theories are applicable to our data.

What is the origin of this doping dependence of the fluctuation strength? The Ginzburg temperature contains the ratio $k_B T_c / H_c^2(0)\xi_0^3$, and furthermore we recall that the scaling means that data taken in different fields are normalized by the ratio of the coherence volume to the magnetic length $(\Phi_0/B)^{1/2}$ (the latter is only determined by the magnetic field and is sample independent). Therefore, the doping dependence of the of the fluctuation regime width must originate from the doping dependence of $H_c(0)$ and $\xi_0^3$. The doping dependence of $\xi_0^3$ can be verified by the doping dependence of the upper critical field $H_{c2}(0)=\Phi_0/(2\pi\xi_0^2)$. The extrapolated $H_{c2}$ values for fields applied perpendicular to the FeAs layers for our samples with x = 0.22, 0.35, 0.50 and 0.60 are 60 T, 173 T, 124 T and 85 T, respectively [18]. The corresponding values for the average in-plane coherence lengths $\xi_{ab}(0)$ are added in Fig. 6 (stars). It can be seen that $\xi_{ab}$ reaches a minimum in the vicinity of optimal doping, which should largely reflect the doping dependence of the coherence volume. A small coherence volume increases the fluctuation range and hence the coherence volume is unlikely the dominant factor determining the doping dependence of the fluctuation strength in $Ba_{1-x}K_xFe_2As_2$. However, $H_c(0)$ (shown as pentagons in Fig. 6), and thus also the condensation energy, shows a peak at optimal doping, which counteracts the shrinking coherence length and thus reduces the fluctuations [30]. In addition, the width of the critical regime is significantly reduced in strong coupling superconductors [31]. The superconducting coupling strength can be estimated from the size of the normalized specific heat jump $\Delta C/\gamma T_c$ [30], (see diamonds in Fig. 6). 6. $\Delta C/\gamma T_c$ shows a peak which is significantly larger than the BCS value $\Delta C/\gamma T_c = 1.43$ at optimal doping, and thus the smaller fluctuation range around optimal doping is likely the result of a larger condensation energy and a stronger coupling strength.

## V. CONCLUDING REMARKS

Our data show that, although critical fluctuation effects are comparatively weak in the family of 122 Fe-based superconductors in zero field [12], they significantly influence the superconducting properties in a wide temperature range of several Kelvins around $T_c(H)$ in applied magnetic fields of a few Tesla. The $T_c$ broadening in magnetic fields obeys the expected scaling predictions of the 3d-XY and the 3d-LLL universality classes. This demonstrates that the broadening is caused by a magnetic field induced finite size effect [2], induced by the magnetic length $(\Phi_0/B)^{1/2}$ associated with the vortex lattice spacing and the confinement of the electrons in low Landau orbits. The vortex melting transition line from a solid vortex phase into a phase incoherent vortex liquid phase [14] follows the same scaling predictions, which demonstrates that the vortex melting is caused by the increasing strength of fluctuations upon approaching the fluctuation regime [14].

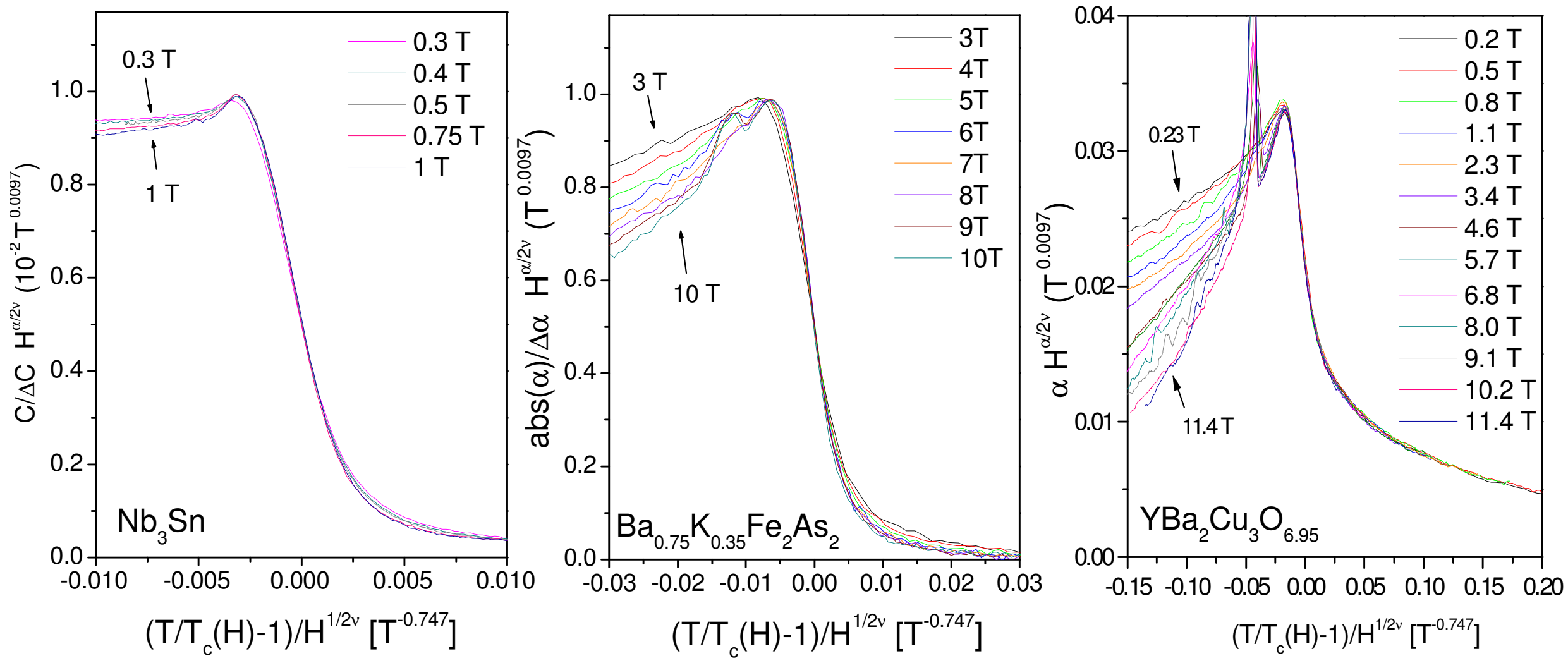

**Figure 7.** Comparison of the 3d-XY scaling plots of the specific heat of the classical superconductor $Nb_3Sn$ [9] and the thermal expansion coefficients of the Fe-based pnictide superconductor $Ba_{0.65}K_{0.35}Fe_2As_2$ and the cuprate high temperature superconductor $YBa_2Cu_3O_{6.95}$ [2]. The additional sharp peaks in the data of $YBa_2Cu_3O_{6.95}$ are irreversible anomalies caused by an underlying vortex glass transition [29].

Upon comparison of $Ba_{1-x}K_xFe_2As_2$ with the cuprate high-temperature superconductor $YBa_2Cu_3O_{7-\delta}$ and the classical superconductor $Nb_3Sn$, we can confirm that the strength of fluctuations in $Ba_{1-x}K_xFe_2As_2$ is intermediate between cuprates and classical superconductors. Fig. 7 shows 3d-XY scaling plots for specific heat data of $Nb_3Sn$ [9] and thermal expansion data of $Ba_{0.65}K_{0.35}Fe_2As_2$ (same as in Fig. 4) and $YBa_2Cu_3O_{6.95}$ [2]. Note the difference in the ranges of the x-axis for the three samples, which illustrates the large variation in the width of the fluctuation regime by more than a factor of 20. In the following, we consider the upper boundary $T_{onset}$ of the *x*-axis scaling variable $[T/T_c(H)\text{-}1]H^{-1/2\nu}$, for which scaling of a visible fluctuation contribution holds, as a rough measure of the strength of fluctuations. Strong fluctuations appear in $YBa_2Cu_3O_{6.95}$ up to at least $[T_{onset}/T_c(H)\text{-}1]H^{-1/2\nu}=0.2$ [2], while weak fluctuations extend in $Nb_3Sn$ up to $[T_{onset}/T_c(H)\text{-}1]^{-1/2\nu}=0.01$ [9]. For optimally doped $Ba_{0.65}K_{0.35}Fe_2As_2$ fluctuation contributions extend up to an intermediate value of $[T_{onset}/T_c(H)\text{-}1]H^{-1/2\nu}=0.02$, and this value increases further in underdoped samples with x = 0.22 up to $[T_{onset}/T_c(H)\text{-}1]H^{-1/2\nu}=0.06$. We have demonstrated that the doping dependence of the width of the fluctuation regime in $Ba_{1-x}K_xFe_2As_2$ is likely driven by the superconducting condensation energy and coupling strength.

ACKNOWLEDGMENTS

This work was supported by grants from the Research Grants Council of the Hong Kong Special Administrative Region, China (603010, SEG_HKUST03, SRFI11SC02).